\documentclass[a4paper,12pt]{article}

\title{Habitat fragmentation: the possibility of a patch disrupting its neighbor}

\author{${}^{1}$Pamplona da Silva D. J. and ${}^{2}$Capelas de Oliveira, E.\\
{\scriptsize ${}^{1}$Instituto de Ci\^encia e Tecnologia - UNIFAL-MG 37715-400 Po\c{c}os de Caldas, MG, Brazil, pamplona@unifal-mg.edu.br}\\
{\scriptsize ${}^{2}$Depto. Matem\'atica Aplicada - Unicamp 13083-859 Campinas, SP, Brazil}}

\usepackage{graphicx}
\usepackage{color}

\begin{document}

\maketitle

\begin{abstract}
\noindent This paper starts from the Fisher-Kolmogorov-Petrovskii-Piskunov equation to model diffusive populations. The main result, according to this model, is that two connected patches in a system do not always contribute to each other. Specifically, inserting a large fragment next to a small one is always positive for life inside the small patch, while inserting a very small patch next to a large one can be negative for life  inside the large fragment. This result, obtained to homogeneously fragmented regions, is possible from the general case expression for the minimum sizes in a system of two patches. This expression by itself is an interesting result, because it allows the study of other characteristics not included in the present work.
\end{abstract}

\section{Introduction}

Habitat fragmentation is a problem for survival of different species living in several environments, from mammals scattered around the globe \cite{Crooks2017} to bacteria in the laboratory \cite{Kenkre2003, Perry2005}. Concerns about the relation between fragment size and population density have also been discussed in the literature \cite{Bowers1997}, concluding that larger and more intact areas are more beneficial for species preservation \cite{Connor2000}, without neglecting small fragments which have a fundamental contribution to biodiversity and species conservation \cite{Lindenmayer2018}, sometimes preventing some of them from extinction \cite{Wintle2018}.

Neither the relevance of intact forests nor the importance of small fragments dismiss the modelling about the minimum patch size \cite{Kenkre2003,Perry2005,Kraenkel2010,PamplonadaSilva2018}, because the fragments considered small can be sufficiently big to maintain a stable population of a given species, once increasing their size does not imply an increase in the population density \cite{Bowers1997}. In this sense, not so recent mathematical models \cite{Fisher1937, Skellam1951} and their successors \cite{PamplonadaSilva2018,Kenkre2008,Ludwig1979} have been found a minimum patch size that enables stable life inside the fragment. These mathematical results can be used in experimental discussions \cite{Perry2005}. Similarly, experimental results are important guides for extending or refining models so that they can describe reality.

From previous studies \cite{PamplonadaSilva2012, PamplonadaSilva2017}, as well as in the literature \cite{Kenkre2008},   in a system of two communicating patches, both need a smaller size to provide life inside them \cite{Kenkre2008, PamplonadaSilva2012, PamplonadaSilva2017}, if compared with an alone patch.The decrease in the minimum size of each fragment comes from the communication between them through a region where it is possible the diffusion of the species in question, but it can not be a definitive address for this population. However, these studies considered only equal fragments, so this behavior becomes doubtful in a system with asymmetric fragments.

In this sense, this work is a natural continuation of previous studies \cite{ PamplonadaSilva2018, PamplonadaSilva2012, PamplonadaSilva2017}, and its main purpose is to answer if the presence of a small fragment is good, bad or indifferent to another fragment near it. In this context, it is understood as close, the patches connected as described bellow. A good test to achieve this goal is to start from the smallest patch (critical size) that alone maintains a stable population inside it and observe how this critical size behaves after inserting an another very small fragment in its neighborhood.

To answer this question, it is need a prediction about how the patch size behaves in a system of two totally asymmetric patches. Mathematically this implies to find an analytic expression for the minimum size, as found in the literature for particular cases \cite{PamplonadaSilva2018, Kenkre2008, PamplonadaSilva2012, PamplonadaSilva2017}. Here, in this paper, the objective is to present the general case.

The paper is disposed as follows: in section \ref{terminology} will present the problem and discuss known techniques and results that will be useful to obtain new results of this work and their comparison with the literature. In section \ref{discussion} it will be presented a brief mathematical discussion indicating the main steps for obtaining an analytical prediction of minimum patch sizes for the general asymmetric case of homogeneous regions. In section \ref{results} will address both mathematical and phenomenological (model-based predictions) results as well as their discussion. Concluding remarks close the paper.

\section{Preliminaries and notation} \label{terminology}

The Fisher-Kolmogorov-Petrovskii-Piskunov (FKPP) equation has been used to model population dynamics from microscopic species as genes \cite{Fisher1937}, bacteria \cite{Kenkre2003, Perry2005} and cells \cite{Takamizawa1997}, to macroscopic one, like spruce budworm \cite{Ludwig1979}, so that its application in the study of diffusive populations is always a valid attempt.

In population dynamics, one feature is the existence of a critical size $L_{c}$, below which a stable population cannot exist within the fragment. This critical size can be used as the minimum size for the preservation of a species, or as the maximum size that guarantees the extinction of this species, i.e. if the patch size $L$ is smaller than $L_{c}$, the species will be extinct from this patch.

The most general form of the one-dimensional FKPP equation found in the literature \cite{Kenkre2003, Skellam1951, Ludwig1979} as a mathematical model for describing a diffusive population is:

\begin{equation}\label{eqfish}
\frac{\partial u}{\partial t}=D\frac{\partial^2 u}{\partial x^2}+a(x)u-bu^{2},
\end{equation}

\noindent where the variables related to population are: $u=u(x, t)$ the density, $a(x)$ the growth rate, $b\geq 0$ the saturation rate or intraspecific competition, and $D$ the dispersion coefficient. The $x$ and $t$ variables represent space and time respectively.

\begin{figure}[ht]
\begin{center}
\includegraphics[width=.35\linewidth]{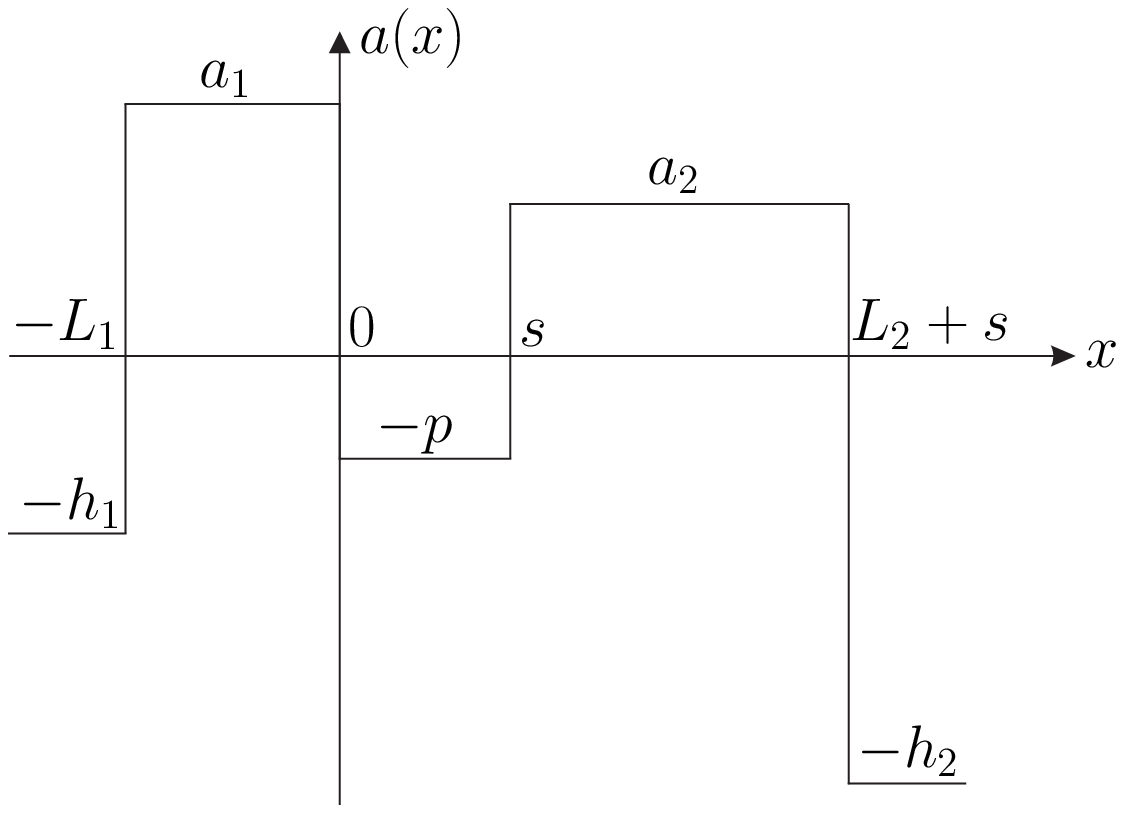} 
\hspace{0.1cm} \includegraphics[width=.25\linewidth]{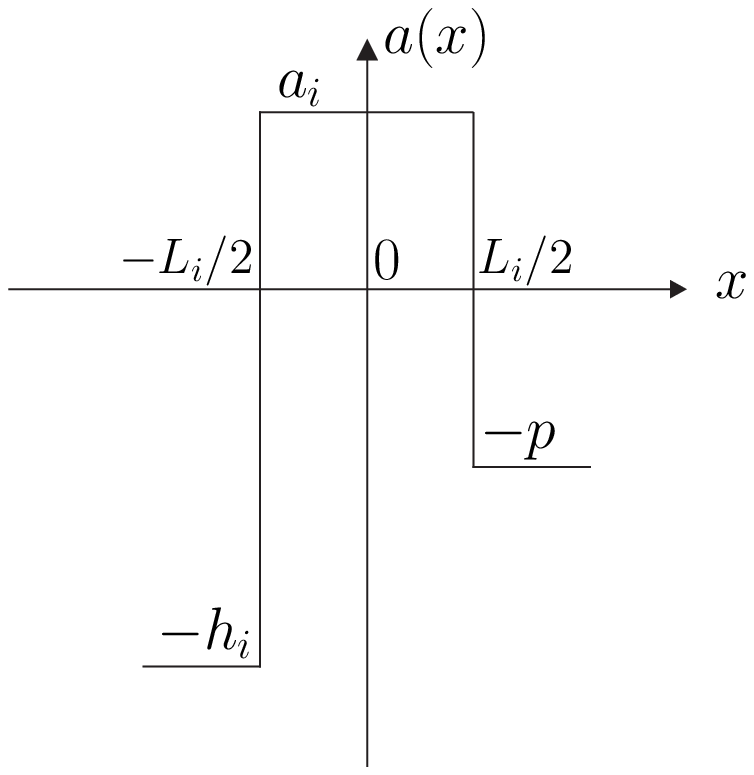}\\
(a) \hspace{4.5cm} (b) 
\caption{\label{duplogeral} \small Representation of growth rate profiles used to describe the general case of (a) a two patches system and (b) a single patch.}
\end{center}
\end{figure}

The $a(x)$ profile is commonly used to represent space fragmentation, because if $a(x)<0, \forall x$ in Eq.(\ref{eqfish}), the population density $u(x,t)$ goes to zero for large times ($t\rightarrow\infty$), what does not happen if $a(x)>0,\forall x$. However, if $a(x)$ is a composition of regions where $a(x)$ is sometimes positive and sometimes negative, it is possible a stable solution $u(x,t)$ that satisfies Eq.(\ref{eqfish}) in the considered domain. A common interpretation is to take over as patches (or fragments), the regions where $a(x)>0$, surrounded by unsuitable regions, where $a(x)<0$, this last will be labeled here as sinks, see Fig.(\ref{duplogeral}).

In the proposed model, the simplest possible fragmentation is to merge homogeneous regions (where $a(x)$ is constant), sometimes favorable, sometimes unfavorable to life. For this type of fragmentation the most general form of a two fragment system is represented by Fig.(\ref{duplogeral}a), which is composed of two patches and three sinks, as identified in the Table \ref{definitions}.

\begin{table}[ht]
\begin{center}
{\footnotesize
\caption{\label{definitions} Definition of regions and their properties.}
\begin{tabular}{|c|c|c|}
\hline  
Short label & Region label & Space region\\ 
\hline
$S1$ & sink 1 & $x<-L_{1}$\\
\hline
$P1$ & patch 1 & $-L_{1}\leq x \leq 0$\\
\hline
$Ss$ & sink s &  $0<x<s$\\
\hline
$P2$ & patch 2 & $s\leq x \leq L_{2}+s$\\
\hline
$S2$ & sink 2 & $x>L_{2}+s$\\
\hline
\end{tabular}}
\end{center}
\end{table}

In Fig.(\ref{duplogeral}b) the general case of one only patch is presented. Its critical size was recently found by Pamplona da Silva \cite{PamplonadaSilva2018} in explicit form:
\begin{equation}\label{eqLsph}
L_{sph}=\sqrt{\frac{D}{a_{i}}}\left\{\arctan{\left(\sqrt{\frac{h_{i}}{a_{i}}}\right)}+\arctan{\left(\sqrt{\frac{p}{a_{i}}}\right)}\right\}.
\end{equation}

From Eq.(\ref{eqLsph}) the critical size for a semi-isolated fragment is easily obtained at the limit $h_{i}\rightarrow\infty$, i.e.,
\begin{equation}\label{eqLspi}
L_{spi}=\sqrt{\frac{D}{a_{i}}}\left\{\frac{\pi}{2}+\arctan{\left(\sqrt{\frac{p}{a_{i}}}\right)}\right\}.
\end{equation}

Fig.(\ref{duplogeral}b) can be obtained from Fig.(\ref{duplogeral}a) taking the limit $s\rightarrow\infty$. The variable $h_{i}$ is used to represent $h_{1}$ and $h_{2}$ and $a_{i}$ to represent $a_{1}$ and $a_{2}$, alluding now to fragment 1 and 2, respectively.

\section{Mathematical discussion}\label{discussion}

To describe a time-varying population it is necessary a model with a temporal dependence and that the population cannot grow infinitely. In this sense, the Eq.(\ref{eqfish}) is minimally qualified to model a diffusive population dynamics at any point ($x,t$) of spacetime. However, as already discussed by Pamplona da Silva et al. \cite{PamplonadaSilva2018, PamplonadaSilva2012, PamplonadaSilva2017} and Kenkre and Kumar \cite{Kenkre2008}, following the ideas of Ludwig et al. \cite{Ludwig1979}, in the limit condition, it is possible to work only with the linear stationary equation, namely
\begin{equation}\label{eqespacial}
D\frac{\partial^2 \Phi}{\partial x^2}+a(x)\Phi=0,
\end{equation}
where $\Phi=\Phi(x)$.

In order to find the critical sizes $L_{1}$ and $L_{2}$ of fragments 1 and 2 respectively, it is discuss the Eq.(\ref{eqespacial}), region by region, (see Table \ref{definitions}), enforcing the continuity of the function and its derivative at the boundaries. Thus, the functions which are solutions of Eq.(\ref{eqespacial}) in five regions, are initially identified:
\begin{equation}\label{PhiI}
\Phi_{S1}(x)=Ae^{k_{I}x} + A_{1}e^{-k_{I}x},
\end{equation}
\begin{equation}\label{PhiII}
\Phi_{P1}(x)=C\sin{(\alpha_{II}x)}+D\cos{(\alpha_{II}x)},
\end{equation}
\begin{equation}\label{PhiIII}
\Phi_{Ss}(x)=Ge^{k_{III}x}+He^{-k_{III}x},
\end{equation}
\begin{equation}\label{PhiIV}
\Phi_{P2}(x)=E\sin{(\alpha_{IV}x)}+F\cos{(\alpha_{IV}x)},
\end{equation}
\begin{equation}\label{PhiV}
\Phi_{S2}(x)=B_{1}e^{k_{V}x}+Be^{-k_{V}x},
\end{equation}
where $k_I^2=h_1/D$, $\alpha_{II}^2=a_{1}/D$, $k_{III}^2=p/D$, $\alpha_{IV}^2=a_{2}/D$ e $k_V^2=h_{2}/D$.

From the boundary conditions $\Phi_{I}(-\infty)=0$ and $\Phi_{V}(\infty)=0$, it is obtained the constants $A=0$ and $B=0$ respectively. With the functions (solutions of Eq.(\ref{eqespacial}) in their respective regions) given by Eq.(\ref{PhiI}), Eq.(\ref{PhiII}), Eq.(\ref{PhiIII}), Eq.(\ref{PhiIV}) and Eq.(\ref{PhiV}), it is enforced continuity at borders: $x=-L_{1}, x=0, x=s$ and $x=L_{2}+s$, respectively obtaining the expressions:
%
%
\begin{equation}\label{contphil1}
Ae^{-k_{I}L_{1}}=-C\sin{(\alpha_{II}L_{1})}+D\cos{(\alpha_{II}L_{1})},
\end{equation}
%
%
\begin{equation}\label{DGH}
D=G+H,
\end{equation}
%
%
\begin{equation}\label{contphis}
Ge^{k_{III}s}+He^{-k_{III}s}=E\sin{(\alpha_{IV}s)}+F\cos{(\alpha_{IV}s)},
\end{equation}
%
%
\begin{equation}\label{contphil2s}
E\sin{\alpha_{IV}(L_{2}+s)}+F\cos{\alpha_{IV}(L_{2}+s)}=Be^{-k_{V}(L_{2}+s)},
\end{equation}
as well as the continuity of the first derivative at the same points provides:
%
%
\begin{equation}\label{contphilinhal1}
k_{I}Ae^{-k_{I}L_{1}}=\alpha_{II}C\cos{(\alpha_{II}L_{1})}+\alpha_{II}D\sin{(\alpha_{II}L_{1})},
\end{equation}
%
%
\begin{equation}\label{CGH}
\alpha_{II}C=k_{III}(G-H),
\end{equation}
%
%
\begin{equation}\label{contphilinhas}
\begin{array}{cc}
k_{III}Ge^{k_{III}s}-k_{III}He^{-k_{III}s}=&\\
&\hspace{-2.4cm}\alpha_{IV}E\cos{(\alpha_{IV}s)}-\alpha_{IV}F\sin{(\alpha_{IV}s)},
\end{array}
\end{equation}
%
%
\begin{equation}\label{contphilinhal2s}
\begin{array}{cc}
\alpha_{IV}E\cos{\alpha_{IV}(L_{2}+s)}-\alpha_{IV}F\sin{\alpha_{IV}(L_{2}+s)}=&\\
&\hspace{-2.4cm}-k_{V}Be^{-k_{V}(L_{2}+s)}.
\end{array}
\end{equation}

To solve the algebraic system composed by Eq.(\ref{contphil1}) to Eq.(\ref{contphilinhal2s}), in order to determine the constants, several paths can be followed to eliminate redundant equations and to obtain a linearly independent system. If it is desired to get exactly and directly the same format shown in this paper, just follow these steps: from Eq.(\ref{contphil1}) and Eq.(\ref{contphilinhal1}), obtain the relation $C=R_ {1}D$ (see Eq.(\ref {where}), below) and replace it in linear combinations of Eq.(\ref{DGH}) and Eq.(\ref{CGH}) resulting in one relation between $G$ and $D$ and other between $H$ and $D$. Similarly, with Eq.(\ref{contphis}), Eq.(\ref{contphil2s}), Eq.(\ref{contphilinhas}) and Eq.(\ref{contphilinhal2s}) find a relation between $D$ e $F$ and other between $H$ and $F$. Finally eliminate the constants $G$ and $H$ by obtaining a system in the variables $D$ and $F$, which in matrix form can be expressed by:
\begin{equation}\label{LIsistem}
\hspace{-0.6cm}\left(\!\!\!\begin{array}{ll} 
\displaystyle  (m_{1}R_{1}+m_{2})e^{-k_{III} s} & 
\displaystyle -(\alpha_{II} R_{1}+k_{III})\\
\\ 
\displaystyle (m_{3}R_{2}+m_{4})e^{k_{III} s} & 
\displaystyle \alpha_{II} R_{1}-k_{III}\\
\end{array}\!\!\!\right)
\!\!\left(\!\!\begin{array}{l}
\displaystyle  F \\
\\ 
\displaystyle  D \\
\end{array}\!\!\right)
\!\!=
\!\!\left(\!\!\begin{array}{l}
\displaystyle  0 \\
\\ 
\displaystyle  0 \\
\end{array}\!\!\right)
\end{equation}
where
\begin{equation}\label{where}
\begin{array}{l}
\vspace{0.15cm}
\displaystyle R_{1}=\frac{k_{I}\cos{(\alpha_{II} L_{1})}-\alpha_{II}\sin{(\alpha_{II} L_{1})}}{\alpha_{II}\cos{(\alpha_{II} L_{1})}+k_{I}\sin{(\alpha_{II} L_{1})}},\\
\vspace{0.15cm}
\displaystyle R_{2}=\frac{k_{V}\cos{(\alpha_{IV} (L_{2}+s))}-\alpha_{IV}\sin{(\alpha_{IV} (L_{2}+s))}}{\alpha_{IV}\cos{(\alpha_{IV} (L_{2}+s))}+k_{V}\sin{(\alpha_{IV} (L_{2}+s))}},\\
m_{1}=-[\alpha_{IV}\cos{(\alpha_{IV} s)}+k_{III} \sin{(\alpha_{IV} s)}],\\
m_{2}=k_{III}\cos{(\alpha_{IV} s)}+\alpha_{IV}\sin{(\alpha_{IV} s)},\\
m_{3}=\alpha_{IV}\cos{(\alpha_{IV} s)}-k_{III}\sin{(\alpha_{IV} s)},\\
m_{4}=k_{III}\cos{(\alpha_{IV} s)}+\alpha_{IV}\sin{(\alpha_{IV} s)}.\\
\end{array}
\end{equation}

The nontrivial solution for the system of Eq.(\ref{LIsistem}) generates the secular equation Eq.(\ref{gen_cond}) which will be presented in the Section \ref{results}.

\section{Results and discussion}\label{results}

This work has a mathematical result which is an interesting result by itself, since it generalizes, in the critical condition, the relation between the problem parameters ($L_{1}$, $L_{2}$, $a_{1}$, $a_{2}$, $h_{1}$, $h_{2}$, $D$, $p$, and $s$). This result arises from the secular equation associated with the system of Eq.(\ref{LIsistem}),
\begin{equation}\label{gen_cond}
\begin{array}{cc}
(m_{1}R_{2}+m_{2})(\alpha_{II} R_{1}-k_{III})e^{-k_{III} s}+&\\
&\hspace{-3.8cm}(m_{3}R_{2}+m_{4})(\alpha_{II} R_{1}+k_{III})e^{k_{III} s}=0.
\end{array}
\end{equation}

Eq.(\ref{gen_cond}) provides many possibilities of analysis, including particular one, which has been a subject of interest in the literature \cite{PamplonadaSilva2018, Skellam1951, Ludwig1979} and is the main phenomenological focus of this paper, namely, the relation between fragment sizes ($L_{1}$ and $L_{2}$), in critical condition. Specifically, how the presence of a patch affects its neighbor critical size, as discussed below.

\begin{figure}[!ht]
\begin{center} 
\includegraphics[width=9.5cm]{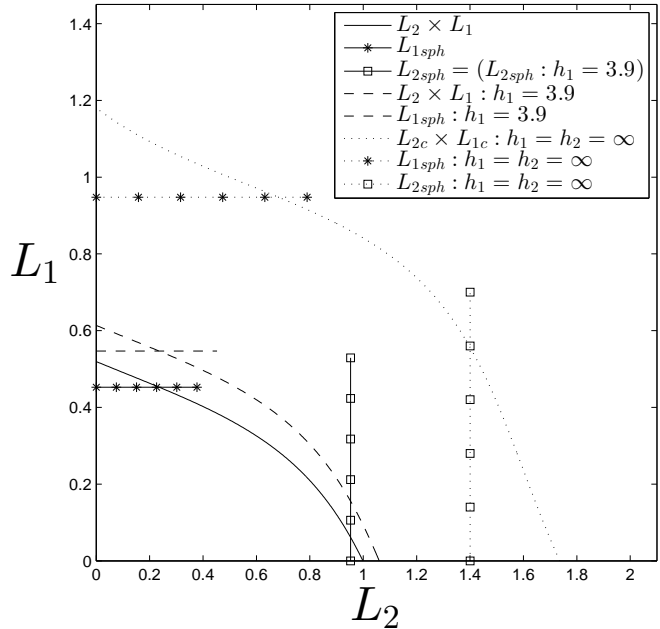} 
\caption{\label{resl1l2ld} \small Critical patch size for parameters $a_{1}=4.8$, $a_{2}=2.5$, $h_{1}=1.7$, $h_{2}=3.9$, $D=1.1$, $p=0.9$ and $s=0.4$, (solid lines), variations in life difficulty conditions inside the sinks, $h_{1}=3.9$ (dashed lines) and $h_{1}=h_{2}=\infty$ (dotted lines).}
\end{center}
\end{figure}

Fig.(\ref{resl1l2ld}) shows (oblique continuous line) the relation between $L_{1}$ and $L_{2}$ for the parameter set, $a_{1}=4.8$, $a_{2}=2.5$, $h_{1}=1.7$, $h_{2}=3.9$, $D=1.1$, $p=0.9$ and $s=0.4$. In this one, it is observed that in the presence of a very small patch 2, the fragment 1 has a critical size ($L_{1}$) larger than if it was alone ($L_{1sph}$), i.e., the presence of patch 2 too small is bad for fragment 1. Similarly, fragment 2 has its critical size increased by the proximity of a very small patch 1, see regions $L_{1}>L_{1sph}$ and $L_{2}>L_{2sph}$.


Further, in Fig.(\ref{resl1l2ld}) the life difficulty condition in sink 1 is increased (dashed lines with the same solid lines parameters except $h_{1}:1.7\rightarrow 3.9$) and not only the patch 1 size need to be larger, but also fragment 2 needs a larger size, even when $L_{1}$ tends to zero. Concluding that for a tiny fragment 1, the patch 2 feels the edge of fragment 1 (sink 1) and does not the patch 1 by itself, independently of life conditions inside patch 1, however good it may be. This argument is confirmed by the dotted lines in Fig.(\ref{resl1l2ld}), which represents the case $h_{1}=h_{2}\rightarrow\infty$, where fragment 1 feels a lot the fragment 2 edge when $L_{2}$ is small and vice-versa. This last result also confirms the results of Pamplona da Silva et al. \cite{PamplonadaSilva2012} and Kenkre and Kumar \cite{Kenkre2008}, namely, the existence of two identical patches with a connection between them (sink $p$) is always positive for both.

\begin{figure}[ht]
\begin{center} 
\includegraphics[width=9.5cm]{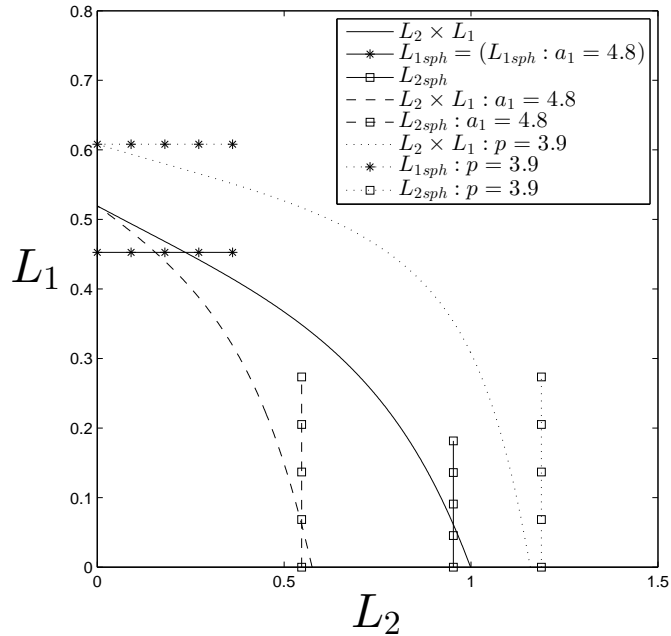}\\
\caption{\label{resl1l2lc} \small Critical patch size for parameters $a_{1}=4.8$, $a_{2}=2.5$, $h_{1}=1.7$, $h_{2}=3.9$, $D=1.1$, $p=0.9$ and $s=0.4$, (solid lines), variations in life condition inside patch 1, $a_{1}=4.8$ (dashed lines) and life difficulty between the patches, $p = 3.9$ (dotted lines).}
\end{center}
\end{figure}

Looking now at the life condition, dashed lines in Fig.(\ref{resl1l2lc}), it has an improvement inside the patch 2, keeping the other parameters. This generates a decrease in the critical size of patch 2, both in presence as in the absence of patch 1. This however does not change the fact that insertion of a very small patch 2 disrupts the patch 1. When patch 2 size goes to zero, its internal improvement ($a_ {2}:2.5\rightarrow 4.8$) is not felt by patch 1, which was already expected, because if patch 2 no longer exists ($L_{2}\rightarrow 0$), it does not matter what is its internal condition ($a_{2}$).

\section{Concluding remarks}\label{conclusion}

The main result of this work appears when there is a minimally sustainable fragment and other very small one is inserted near it. In this configuration the interaction with this new fragment is negative for the original one. The original fragment, which alone could hold a stable population of a given species inside itself, now needs a larger size to be habitable by the same species due to the proximity of a second very small patch. Thus, the patch size is a key feature to life existence in the coexistence of two patches.

The result just presented was only possible on account of the analytic expression, Eq.(\ref{gen_cond}), for the general case of the critical sizes for two patches with homogeneous conditions inside and outside them. Spatial heterogeneities occur only abruptly and at the fragments borders. From this expression, only one parameter was explored, and there are still possibilities of this result to contribute to future works related to this theme.

As mentioned earlier \cite{Kenkre2008, PamplonadaSilva2012}, the interaction between two identical patches connected by a sink, as long as it is not impenetrable ($p<\infty$), is always beneficial to both, because they can be smaller than they were alone. In this sense, this work corroborates the literature results and expands them, because the fragments do not have to be identical for this contribution to happen. They only can not be too small.

The possibility of a patch (1 or 2) disturbing its neighbor disappears (regardless of its size) if the life difficulty in the adjacent sink ($h_{1}$ or $h_{2}$) is equal to or smaller than condition of the sink between them ($p$). Explicitly, if $h_{1}\leq p$ inserting patch 1 will always be beneficial to patch 2, just as if $h_{2}\leq p$ inserting patch 2 will always be beneficial to patch 1. This effect can be seen in Fig.(\ref{resl1l2lc}) by comparing $(L_{2}\times L_{1}:p=3.9)$ with $(L_{1sph}:p=3.9)$ and $(L_{2sph}:p=3.9)$.

Finally, the results here obtained are in agreement with the literature \cite{Bowers1997, Connor2000}, because in very small fragments, or very fragmented regions, the species density is smaller, since many of them can extinct themselves by being in a tiny patch or taking refuge in a fragment that is close to a small fragment.

\section*{Acknowledgments}
The author (DJPS) thanks the ``Coordenação de Aperfeiçoamento de Pessoal de Nível Superior'' (CAPES) and the ``Departamento de Matemática Aplicada'' (DMA-UNICAMP).

\bibliographystyle{ieeetr}
\bibliography{pdcg_v2}

\end{document}